\documentclass[a4paper,fleqn,usenatbib,useAMS]{aa}

\usepackage[varg]{txfonts}
\usepackage{graphicx}	
\usepackage{amsmath}	
\usepackage{amssymb}	
\usepackage{multicol}

\usepackage{bm}		
\usepackage{natbib,hyperref,ifthen} 
\usepackage{hyperref}
\usepackage{etoolbox}
\hypersetup{
 colorlinks=true,
 citecolor=blue}
\usepackage{color}   

 \renewcommand\appendix{\par
  \setcounter{section}{0}
  \setcounter{subsection}{0}
  \setcounter{figure}{0}
  \setcounter{table}{0}
  \renewcommand\thesection{ \Alph{section}}
  \renewcommand\thefigure{\Alph{section}\arabic{figure}}
  \renewcommand\thetable{\Alph{section}\arabic{table}}
}

\newcommand{\kms}{\,km\,s$^{-1}$}
\newcommand{\hi}{H{\sc i}\ } 

\usepackage[T1]{fontenc}
\usepackage{ae,aecompl}

\begin{document}

   \title{Exploring the GalMer database: bar properties and non-circular motions}
   \titlerunning{Exploring the GalMer database}

   \author{T. H. Randriamampandry
          \inst{1}
          \and
          N. Deg\inst{1}
          \and
          C. Carignan\inst{1,2}
          \and
           F. Combes\inst{3}
          \and
          K. Spekkens\inst{4}
          }

   \institute{Department of Astronomy, University of Cape Town, Private Bag X3, Rondebosch 7701, South Africa\\
              \email{tokyr@ast.uct.ac.za}
         \and
             Observatoire d'Astrophysique de l'Universit\'e de Ouagadougou (ODAUO), BP 7021, Ouagadougou 03, Burkina Faso           
           \and
             LERMA, Observatoire de Paris, College de France, PSL, CNRS, Sorbonne Univ., UPMC, F-75014, Paris, France 
           \and
             Department of Physics, Royal Military College of Canada, PO Box 17000, Station Forces, Kingston, ON K7K 7B4, Canada
             }

   \date{Received ......; accepted .......}

  \abstract
   {We use Tree-SPH simulations from the GalMer database by Chilingarian et al. to characterize and quantify the non-circular motions induced by the presence of bar-like structures on the observed rotation curve of barred galaxies derived from empirical models of their line-of-sight velocity maps. The GalMer database consists of SPH simulations of galaxies spanning a wide range of morphological types and sizes.}
   {The aim is to compare the intrinsic velocities and bar properties from the simulations with those derived from pseudo-observations. This allows us to estimate the amount of non-circularity and to test the various methods used to derive the bar properties and rotation curves.}
   {The intrinsic velocities in the simulations are calculated from the gravitational forces whereas the observed rotation velocities are derived by applying the ROTCUR and DiskFit algorithms to well-resolved observations of intermediate-inclination, strongly barred galaxies.}
   {Our results confirm that the tilted ring method implemented in {\sc ROTCUR} systematically underestimates/overestimates the rotational velocities by up to 40 percent in the inner part of the galaxy when the bar is aligned with one of the symmetry axes for all the models. For the {\sc DiskFit} analysis, we find that it produces unrealistic values for all the models used in this work when the bar is within $\sim$10 degrees of the major or minor axis. }
   {}


   \maketitle
   
\hfill
\section{Introduction}
Bars are common features in disc galaxies, with nearly two thirds of nearby galaxies showing strong or weak bars (eg. \citealt{1991rc3..book.....D, 2000AJ....119..536E, 2000ApJ...529...93K,  2007ApJ...659.1176M, 2007ApJ...657..790M, 2008ASPC..396..351B, 2009A&A...495..491A, 2009ASPC..419..138M, 2010ApJ...711L..61M, 2011MNRAS.411.2026M}).  These bars are dynamically important as they drive secular evolution by moving gas towards the galactic center (\citealt{2013MNRAS.429.1949A,1994ApJ...424...84H,1993ApJ...414..474S}) and transfer angular momentum throughout the galaxy (\citealt{1972MNRAS.157....1L,1985MNRAS.213..451W,2002MNRAS.330...35A, 2007ApJ...659.1176M}); thereby changing the kinematics of the gas and the properties of the galaxy in general.

Of particular importance is the effect of the bar on the velocity maps and rotation curves.  Typically, the gas is assumed to be moving on circular orbits and its observed velocity is used to determine the mass distribution of galaxies (\citealt{de-Blok:2008oq, 2014MNRAS.439.2132R}). This approximation works 
 quite well for unbarred galaxies since the deviations from circular motion are small \citep{1978PhDT.......195B} but, in barred galaxies, the gas flows along the bar. \cite{1999ApJ...526...97R} using CO(1-0) emission line observations found that the molecular gas in the inner part of barred galaxies is moving along the bar instead of in circular orbits. These flows are non-circular in nature and strongly affect rotation curves derived under the assumption of axisymmetry.  Results from \hi observations have also confirmed that the gas streaming along the bar produces large scale non-circular flows \citep{1978PhDT.......195B} on the observed rotation velocities.  The precise effect of non-circular flows on the observed rotation curves depends on the orientation of the bar relative to the major axis of the observed disk (see e.g. \citealt{2007ApJ...657..773V, 2007ApJ...664..204S, 2008MNRAS.385..553D, 2015MNRAS.454.3743R}).
 
N-body/hydrodynamic simulations are ideal to characterize and quantify these non-circular flows since the exact gravitational potential due to the mass is known for all the snapshots.
In this work, a suite of N-Body simulations is used to quantify the size of the observed deviations and compare them to the intrinsic non-circularity found in the simulations.  We have selected snapshots of isolated galaxies in the GalMer \citep{2010A&A...518A..61C} database that have the strongest bars in order to study the largest possible effects of the bar.  In addition to the characterization and quantization of the non-circularity, we also test the ability of the DiskFit algorithm \citep{2007ApJ...664..204S} to recover the circular and non-circular components of the gaseous 	motion based on the observed velocity map.

The paper is organized as follows: a brief description of the GalMer database is given in Sect. \ref{gal}. In Sect. \ref{prop}, we describe the method used to measure the bar properties. In Sect. \ref{ncm}, we present the method used to characterize and quantify the non-circular motions, and the results are discussed in Sect. \ref{dis}.   Finally, in Sect. \ref{sum}, we summarize the main results and highlight some possible future works.

\section{The GalMer database}
\label{gal}
\begin{figure*}
 \begin{center}$
  \begin{array}{cc}
  \includegraphics[width=170mm]{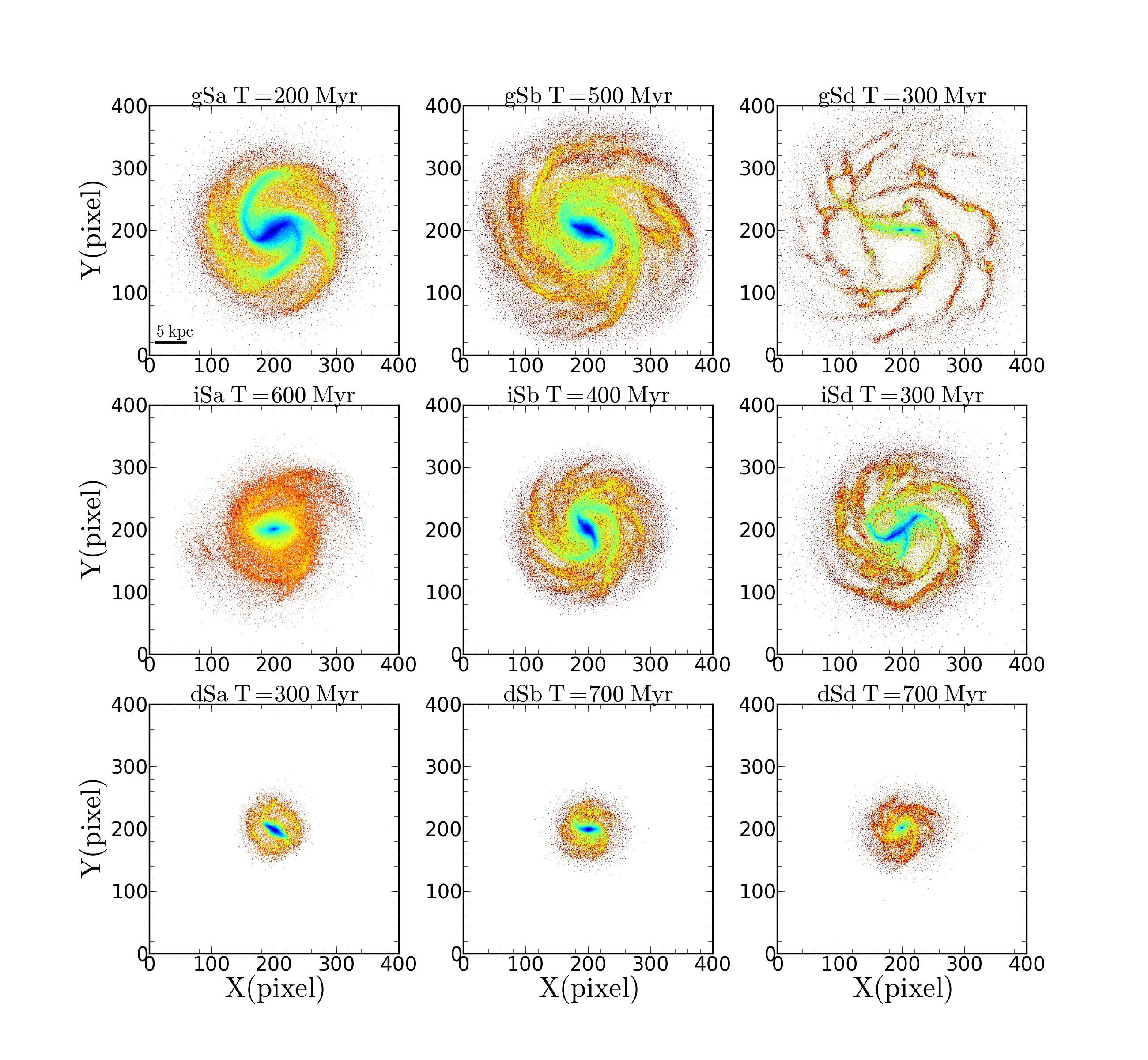}\\
       \end{array}$
    \end{center}
  \caption{Gas and stellar distribution maps for the snapshots with the strongest bars as defined by their bar strengths and lengths. The giants Sa, Sb and Sd are shown on the top panels, the intermediate or medium mass on the middle panels and the dwarf or low mass spirals on the bottom panels. The pixel scale is 0.1 kpc using a zoom of 10 on the online tools (see \url{http://galmer.obspm.fr/}). }
 \label{fignew}
\end{figure*}

The GalMer database \citep{2010A&A...518A..61C} consists of a large set of Tree-SPH simulations of  galaxies with different morphological types and sizes.
The main objective of the GalMer project  is to study galaxy evolution through mergers. It can also be used to study isolated galaxies by choosing the second galaxy pair as "none". 
The simulation results are publicly available and accessible online at \url{http://galmer.obspm.fr/}. The simulations are designed to cover a wide range of morphological types from giant ellipticals to dwarf spirals. 
The initial parameters of the GalMer models used are given in Table 1. 
The models used in this study are: the giant spirals (gSa, gSb and gSd) which are massive and large in size, the intermediate spirals (iSa, iSb and iSd) with medium sizes and masses and the dwarf spirals (dSa, dSb and dSd) with low masses and small sizes. The models are classified as Sa, Sb and Sd galaxies depending on the importance of the bulge component and of the mass fraction (see Table 1). The Sa models have prominent bulge and lower gas fraction similar to early type spirals, the Sb models have smaller bulge but larger gas fraction than the Sa model and the Sd models are bulge-less with a gas fraction similar to late type spirals (See \citealt{2010A&A...518A..61C} for details). The snapshots with the strongest bars are shown in Fig. \ref{fignew} for all nine models used in this work. The giants Sa, Sb and Sd are shown in the top panels, the intermediate or medium mass spirals in the middle panels and the dwarfs or low mass spiral in the bottom panels.

The GalMer database uses a go-on-the-fly algorithm which enables the user to obtain the masses, the velocity maps and other properties for every snapshot. 
 The radial velocities maps are obtained by projecting the velocity components of each particle into a line-of sight (LOS) radial velocity assuming an infinite distance. The pixel size depends on the zoom that is chosen by the user on the online tool.  
 The LOS radial velocity is given by:
  \begin{equation}
V_{r} = V_{X}cos \varphi cos \theta + V_{Y}sin \varphi cos \theta + V_{Z}sin \theta
\end{equation}

where { \it V$_{X}$, V$_{Y}$} and { \it V$_{Z}$} are the velocity components for each particles, $\varphi$ and $\theta$ are the azimuthal and polar angles respectively \citep{2010A&A...518A..61C}. 

The GalMer database does not provide the observed uncertainties for the mock velocity fields.

\begin{table*}
 \caption{Initial parameters for the GALMER simulation from  \citet{2010A&A...518A..61C}. The first column is the model, the second to fifth are the masses of the different components, followed by their scale lengths, and the last three columns are the number of particles for the stars, the gas and the dark matter halo components.}
 \label{tab:anysymbols}
 \tiny
 \begin{center}
 \begin{tabular}{llllllllllll}
  \hline\\
  &M$_{disc}$&M$_{bulge}$&M$_{gas}$&M$_{halo}$ &  a$_{d}$ & r$_{b}$ &a$_{g}$ & r$_{h}$&N. Stars&N. Gas&N. DM\\\\
  Model &10$^{9}$M$_{\sun}$&10$^{9}$M$_{\sun}$&10$^{9}$M$_{\sun}$&10$^{9}$M$_{\sun}$&  kpc& kpc &kpc &  kpc&&&\\\\
1 &2&3&4&5&6&7 &8&9&10&11&12\\\\
  \hline\\
gSa&92.0&23.0&9.4&115.0&4.0&2.0&5.0&10.0&80 000&240 000&160 000\\\\  gSb&46.0&11.5&9.4&172.5&5.0&1.0&6.0&12.0&160 000&160 000&160 000\\\\[2pt]
gSd&57.5&-&7.5&172.5&6.0& -&7.0&15.0&60 000&20 000&40 000\\\\[2pt]
iSa&46.0&11.5&4.6&57.5&2.8&1.4&3.5&7.0&40 000&120 000&80 000\\\\[2pt]
iSb&23.0&5.7&4.6&86.2&3.5&0.7&4.2&8.5&80 000&80 000&80 000\\\\[2pt]
iSd &28.7&-&8.6&86.2&4.2&-&5.0&10.6&120 000&40 000&80 000\\\\[2pt]
 dSa&9.2&2.3&0.9&11.5&1.3&0.3&1.6&3.8&8 000&24 000&16 000\\\\[2pt]
dSb& 4.6&1.2&0.9&17.2&1.6&0.3&1.9&3.8&16 000&16 000&16 000\\\\[2pt]
dSd &5.7&- &1.7&17.2&1.9&-&2.2&4.7&24 000&8 000&16 000\\\\[2pt]
 \hline

 \end{tabular}
   \end{center}
\end{table*}

\section{Bar properties}
\label{prop}

There are several methods used to determine the strength and size of a bar in the literature. The two most commonly used methods are ellipse fitting and Fourier decomposition. Ellipse fitting is often used for observed galaxies, while Fourier decomposition is typically used for simulations.

Ellipse fitting consists of finding the set of iso-density (surface brightness) ellipses.  The ellipticity of each ellipse is then used as a proxy for the bar strength as a function of radius. The largest value of the ellipticity, $\epsilon_{max}$, is used to describe the overall bar strength while the bar length is taken to be the location of rapid changes in either the bar ellipticity of the surface brightness, or defined as the radius where there is a significant change of the position angle of the ellipses (eg. \citealt{2016A&A...587A.160D}).

In Fourier decomposition, the image or density maps are modelled as a Fourier series  \citep{2010ApJ...715L..56S}:
\begin{equation}
 a_{m} = \frac{1}{\pi}\int_{0}^{2\pi} \Sigma (r,\theta) cos(m\theta) d\theta
\end{equation}

\begin{equation}
 b_{m} = \frac{1}{\pi}\int_{0}^{2\pi} \Sigma (r,\theta) sin(m\theta) d\theta
\end{equation}

The $m=0$ and $m=2$ components are calculated.  Then the total second moment, $A_{2}$, is calculated as:

\begin{equation}
 A_{2}(r) = \sqrt{a_{2}^{2}(r) + b_{2}^{2}(r)}
\end{equation}

The phase is found by :

\begin{equation}
 \Phi_{2}(r) = arctan\left(\frac{a_{2}(r)}{ b_{2}(r)}\right)
\end{equation}

The total bar strength is given by the $A_{2}/A_{0}$ component and the bar length is found by either a large dip in surface brightness or a large change of phase (\citealt{2004MNRAS.355.1251L, 2005MNRAS.362.1319L}).
This method has been applied to determine the bar length using the mass distributions of N-body  simulations (e.g. \citealt{2000ApJ...543..704D}), and using images obtained from observations (e.g. \citealt{1998AJ....116.2136A, 2003MNRAS.338..465A}).

\section{Characterization of non-circular motions}
\label{ncm}

\subsection{Different forms of non-circular motions}

 The amplitude of non-circular motions varies between galaxies. These motions complicate the study of the dynamics of gas in disc galaxies and diminish the usefulness of the rotation curves derived assuming axisymmetry. These motions are usually less than 20 \kms\ in magnitude \citep{1978PhDT.......195B} and vary in azimuthal phase relative to the disk major axis so that the overall impact on rotation curves is typically small. 
Other large scale deviations from circular motion can be divided into three types.
The first is a kinematic warp: in this case, the PA and inclination changes with radius(see \citealt{1993ApJ...416...74C, 2007A&A...468..903J}). The commonly used package ROTCUR \citep{1989A&A...223...47B} allows these parameters to vary as a function of radius which make it ideal to deal with this type of non-circular motion. The second type is  large scale asymmetries due to tidal interactions with another galaxy. This produces an unusual shape of the outer part of the \hi\ distribution that typically requires detailed simulation to decipher (eg \citealt{1983MNRAS.203.1253G, 2015ApJ...807...73O}).
The third type, which we are interested in, is the non-circular motion induced by a bar-like structure which is also known as oval or bar distortions (eg. \citealt{2007ApJ...664..204S}). In this case, the deviation from circular flows is due to the gas streaming along the bar.

\subsection{Non-circular motions within the simulations}

The true rotation velocities (expected rotation curve) at a given radius can be obtained from the gravitational potential and its derivatives in the simulation by computing the average forces. 
The expected circular velocity is given by:
\begin{equation}
<V_{expected}^{2}>=r<\frac{\partial \Phi}{\partial r}> = r <F_{r}>
\end{equation}
where $F_{r}$ is the radial force from the particles calculated azimuthally on a grid and $\Phi$ is the gravitational potential . A grid of concentric rings divided into angular bins is used in order to increase the force calculation speed. 

The intrinsic non-circular components of the gas velocities can be determined a number of ways.  A simple approach is to bin the gas particles in radial and angular bins and calculate the differences between the flow within a cell and the expected RC.  These can then be used to calculate the rms for the tangential and radial velocities as a function of radius. 

A slightly more elegant approach is to calculate the $m=0$ and $m=2$ Fourier moments for both the tangential and radial velocities.  This allows a comparison between the strength and orientation of the velocity moments and the mass moments.  It is also particularly useful because the DiskFit algorithm (see Sec. 4.3.2), breaks down the observed velocity map into non-circular radial and tangential components that are similar to the intrinsic Fourier moments.

The Fourier components are given by:

\begin{equation}
V_{t}(r,\theta)=A_{0,t}(r) +  \sum_{m=1}^{\infty}A_{m,t}(r)cos[m\theta + \theta_{m,t}(r)]
\end{equation}

\begin{equation}
V_{r}(r,\theta)=A_{0,r}(r) +  \sum_{m=1}^{\infty}A_{m,r}(r)cos[m\theta + \theta_{m,r}(r)]
\end{equation}

where { \it $A_{m,t}(r)$} and {\it $A_{m,r}(r)$} are the tangential and radial Fourier m$^{th}$ velocity moments respectively, $\theta$, $\theta_{m,t}(r)$ and $\theta_{m,r}(r)$ are the angular phases. 

\begin{figure*}
 \begin{center}$
  \begin{array}{cc}
  \includegraphics[width=180mm]{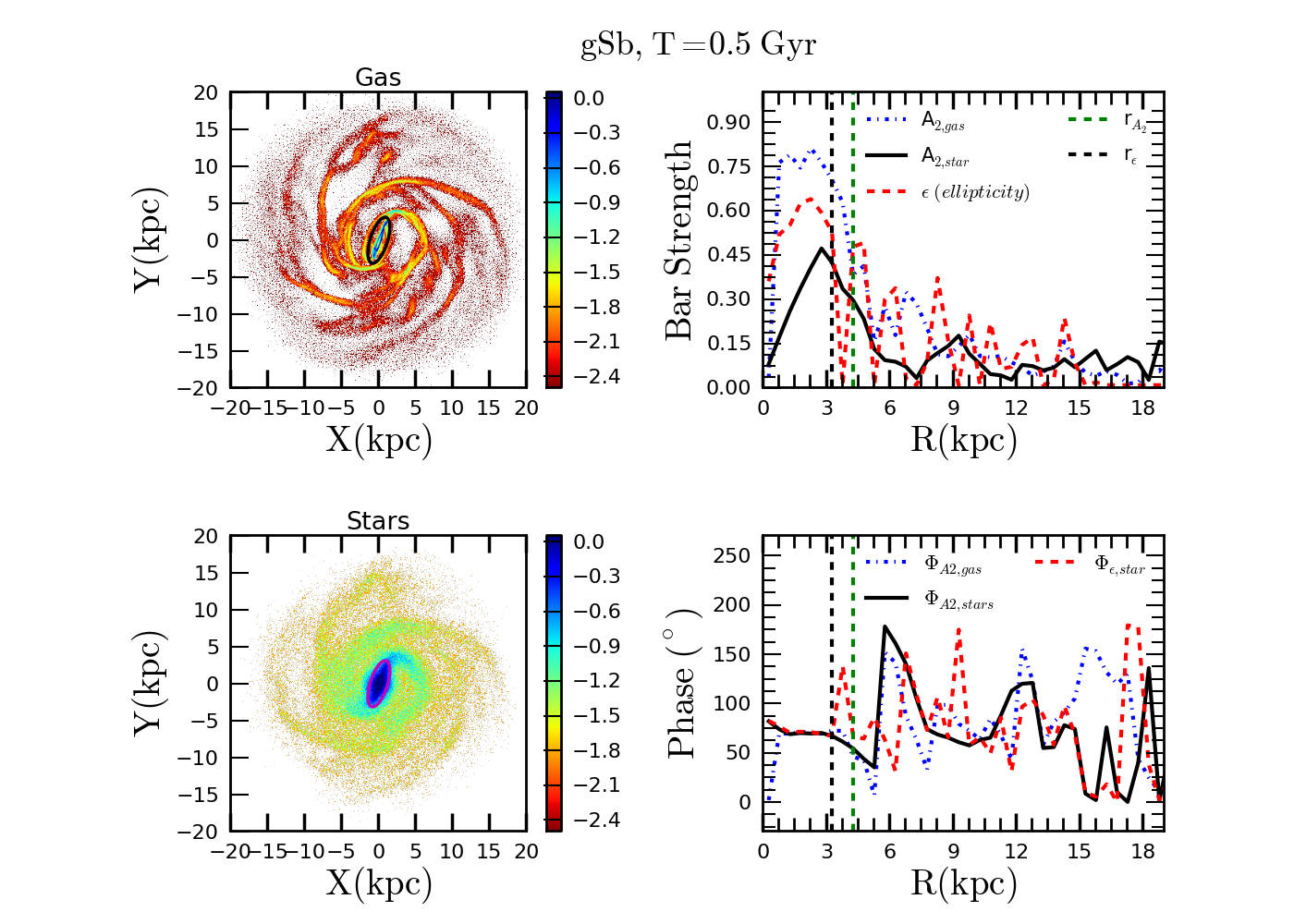}\\
       \end{array}$
    \end{center}
  \caption{Example of bar properties calculation for the gSb model. The gas density map in arbitrary unit is shown in the top left panel and the stellar density map  in the bottom panel. Comparison between the radial variation of the Fourier amplitudes A$_{2}$ derived from the gas density map (dotted blue line), the stellar density map  (continuous black line) and the ellipticity obtained by fitting ellipses to the stellar map (dashed red line) are shown in the top right panel.  The phases are shown in the bottom right panel, lines and symbols are the same in both panels. The vertical dashed green line is the bar length estimated from the Fourier amplitudes, r$_{A2}$, and the vertical black dashed line is the bar length estimated from the ellipticity, r$_{\epsilon}$. The ellipse in the center of each map shows where the bar are located.}
 \label{fig1}
\end{figure*}

\subsection{Non-circular motions from mock observations}

\subsubsection{\rm Tilted ring analysis}
Generally, the observed velocities are described as
 
\begin{equation}
V_{obs} = V_{sys} + V_{c}(r)sin(i) 
\end{equation}

where $V_{sys}$ is the systemic velocity and  $i$ is the inclination angle.  
ROTCUR \citep{1989A&A...223...47B} is widely used to extract $V_{c}$ from two dimensional velocity maps.

The harmonic decomposition, pioneered by \cite{1994ApJ...436..642F}, has been implemented into the GIPSY task RESWRI. 
This task is ideal for small scale departure from circular motions but not for large non-circular flows such as those induced by a bar. We do not include higher order harmonic components in the tilted ring analysis carried out in this work.

\subsubsection{\rm DiskFit analysis}

\cite{2007ApJ...664..204S} adopted a different approach when dealing with large scale non-circular motions induced by bars. Their method is based on modeling the tangential and radial components of the velocity at a given point as a Fourier series and by assuming that the non-circular flow is caused by a bar and that the non-axisymmetric distortion can be described by the m=2 radial and tangential    Fourier moment. 

The DiskFit algorithm is designed to account for non-circular flows such as bi-symmetric distortion, kinematic warp and lopsidedness (see \citealt{2007ApJ...664..204S} for details).  

For a bi-symmetric flow from a bar the velocity map is modeled using:

\begin{equation}
\begin{split}
 V_{model} = V_{sys} + sin(i)[V_{t}cos(\theta) - V_{2,t}cos(2\theta_{b}) cos(\theta) \\\\
 - V_{2,r}sin(2\theta_{b}) sin(\theta)
\end{split}
\end{equation}
where $V_{t}$ is the circular velocity, $V_{2,t}$ and $V_{2,r}$ the amplitudes of the tangential and radial component of the noncircular motions for a bi-symmetric flow model, and $\theta$ and $\theta_{b}$ are the angle in the disk plane relative to the projected major axis and angle relative to the bar axis respectively ( see \citealt{2007ApJ...664..204S}). 

DiskFit uses a bootstrap method to estimate the uncertainties on the parameters using $\chi^{2}$ minimization technique:

\begin{equation}
 \chi^{2} = \sum^{N}_{n=1}(\frac{V_{obs}(x,y)-\sum^{k}_{k=1}\omega_{k,n}V_{k}}{\sigma_{n}})^{2}
\end{equation}

where $V_{\rm obs}$(x,y) is the observed velocity at the position (x,y) on the sky, $\sigma_{n}$ is the uncertainty, $\omega_{k,n}$ is a weighting function which includes the trigonometric factors and also defines an interpolation scheme for the projected model \citep{2007ApJ...664..204S, 2010MNRAS.404.1733S}.

\section{Discussion of the results}
\label{dis}
The bar properties and the effect of the non-circular motion on the rotation curves are discussed in this section. The bar properties (bar strength and bar radius) obtained directly from the simulations and from the images are compared in section \ref{bar}.
In order to see the effect of the bar more clearly, we select the snapshot from each of the nine isolated galaxy runs that has the strongest bar (see Fig \ref{fignew}).  We make mock observations of each snapshot with different bar orientations at a constant inclination of 60 degrees and a pixel size of 0.1 kpc.  Each snapshot was then analyzed using both ROTCUR and DiskFit.  DiskFit returns both the inferred  circular and non-circular
motions, while ROTCUR only returns the circular component.  
Therefore, we define the non-circular motions from ROTCUR as:
\begin{equation}
\rm V_{NCM} = V_{ROTCUR} - V_{expected}
\end{equation}
The ROTCUR analysis is discussed in section \ref{rotc} and the DiskFit analysis in section \ref{disk}. 
\subsection{Bar properties: simulation vs observations}
\label{bar}
An example of the simulation analysis is shown in Fig. \ref{fig1}. The left-hand panels show the gas (upper) and stellar (lower) surface densities for the gSb snapshot. The right-hand panels show the intrinsic bar strengths and orientations using the Fourier and ellipse fitting methods.  There {\bf is} a number of interesting things to note in Fig. \ref{fig1}.  Firstly, the gas and stellar bars are different. The gas bar is significantly stronger, but they both have similar bar lengths.  This is most likely due to the gas' increased sensitivity to the bar's perturbations because of its lower velocity dispersion. Secondly, the ellipse fitting analysis gives a shorter bar. This is due to the Fourier method including similarly oriented spiral structure at larger radii.  The ellipse method more accurately determines the switch from the bar to the spiral structure. \cite{2009A&A...495..491A} argued that the ellipse fitting method could under-estimate the actual bar length, and the difference between the bar length estimated from the Fourier and ellipse fitting technique could also depend on the bar surface brightness profile. The shorter bar obtained from the ellipse fitting technique has also been noticed by other authors (e.g \citealt{2005MNRAS.364..283E, 2006A&A...452...97M}).  

\begin{figure}
 \begin{center}$
  \begin{array}{cc}
  \includegraphics[width=90mm]{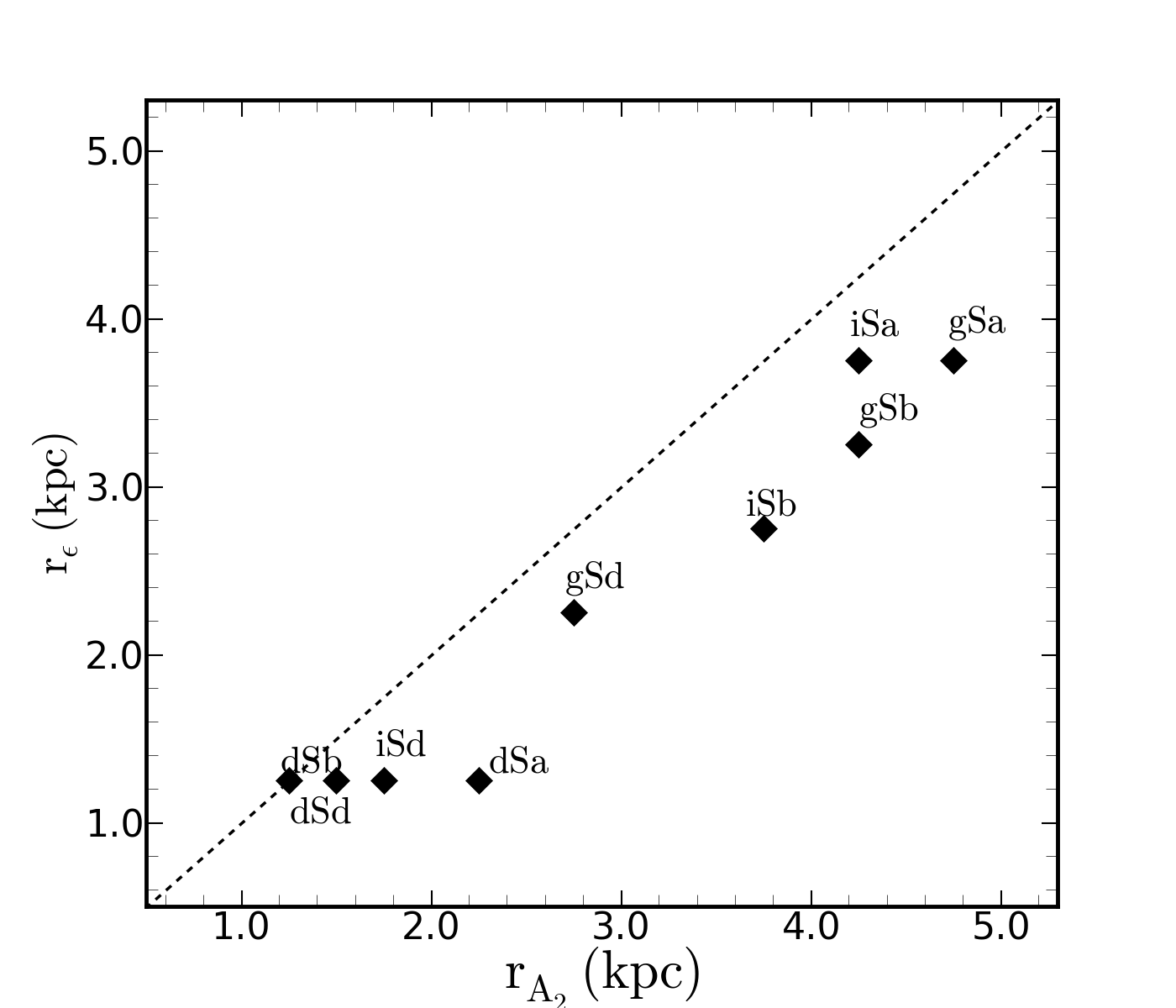}\\
      \end{array}$
    \end{center}
  \caption{Comparison between the intrinsic bar radii obtained using the Fourier decomposition and the ellipses fitting methods estimated from the simulation. The dashed line is the one to one line.}
 \label{fig2}
\end{figure}

In addition, the value of $\epsilon$ is larger than $A_{2}/A_{0}$ for the  stars across the extent of the bar. However, this does not indicate a disagreement regarding the bar strength, because the two quantities are determined using radically different methods. The Fourier method draws concentric circular rings and calculates the Fourier moments, while the ellipse fitting method attempts to draw iso-density ellipses.
 Fig. \ref{fig2} shows comparison between the bar radii estimated from the ellipse fitting to those obtained from the Fourier decomposition for all the models. The bar radii estimated from the ellipse fitting are shorter than those obtained from the Fourier decomposition as illustrated in the example in Fig. \ref{fig1}. 
 
\begin{figure*}
 \begin{center}$
  \begin{array}{cc}
  \includegraphics[width=170mm]{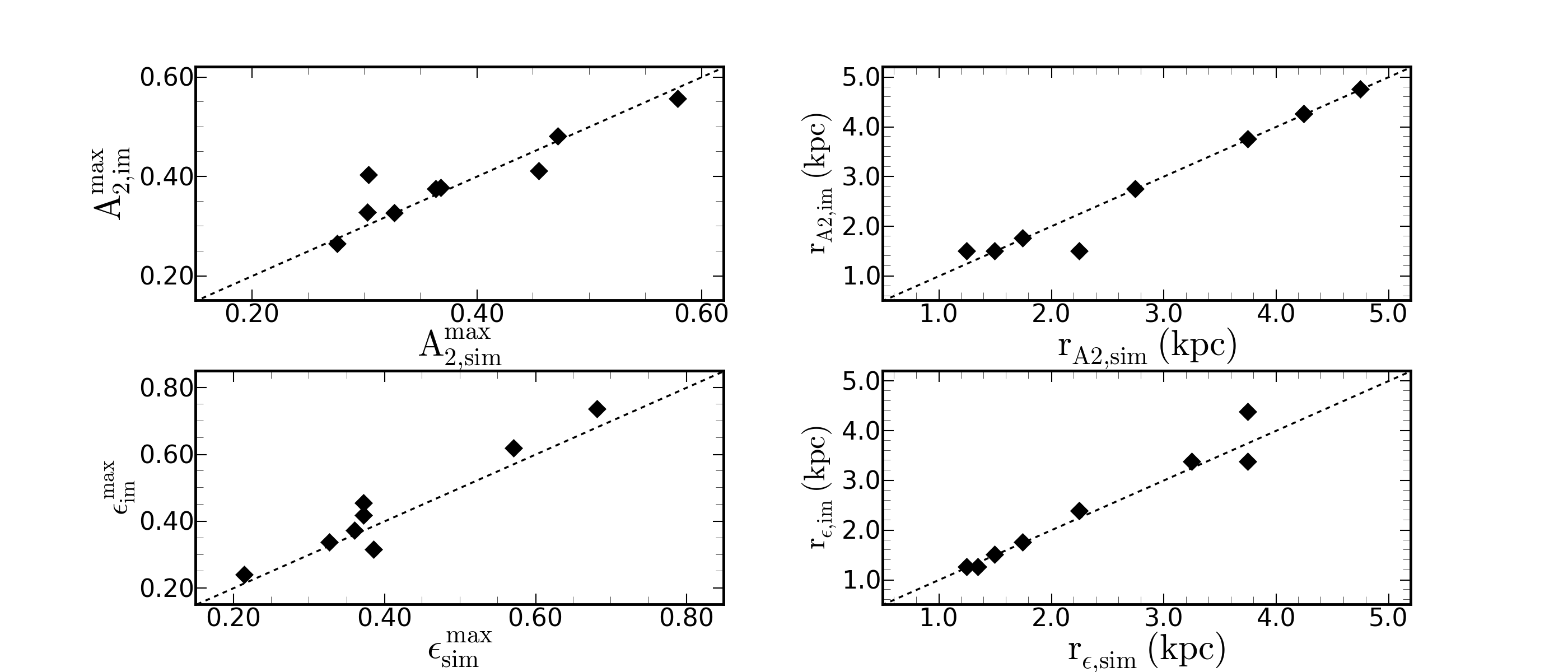}\\
      \end{array}$
    \end{center}
  \caption{Comparison between the bar properties estimated within the simulation to those from images obtained from pseudo-observations. The maximum Fourier m=2 amplitude estimated within the simulation is plotted against those obtained from images on the top left panel. Comparison between the maximum ellipticity measured within the simulation and from images is shown on the bottom left panel. The right panels show comparison between the bar radius estimated within the simulation with those measured from images, the bar radius obtained using the Fourier method is on the top panel and those obtained using the ellipse fitting technique on the bottom panel. }
 \label{fig3}
\end{figure*}

 Fig. \ref{fig3} show{\bf s} comparisons between the simulated and observed bar strengths and radii using the nine snapshots.  This figure shows that the measured bar properties estimated within the simulation are similar to those obtained from pseudo-observations using stellar images.

 \newpage
\subsection{Non-circular motions}

\begin{figure*}
 \begin{center}$
  \begin{array}{cc}
  \includegraphics[width=175mm]{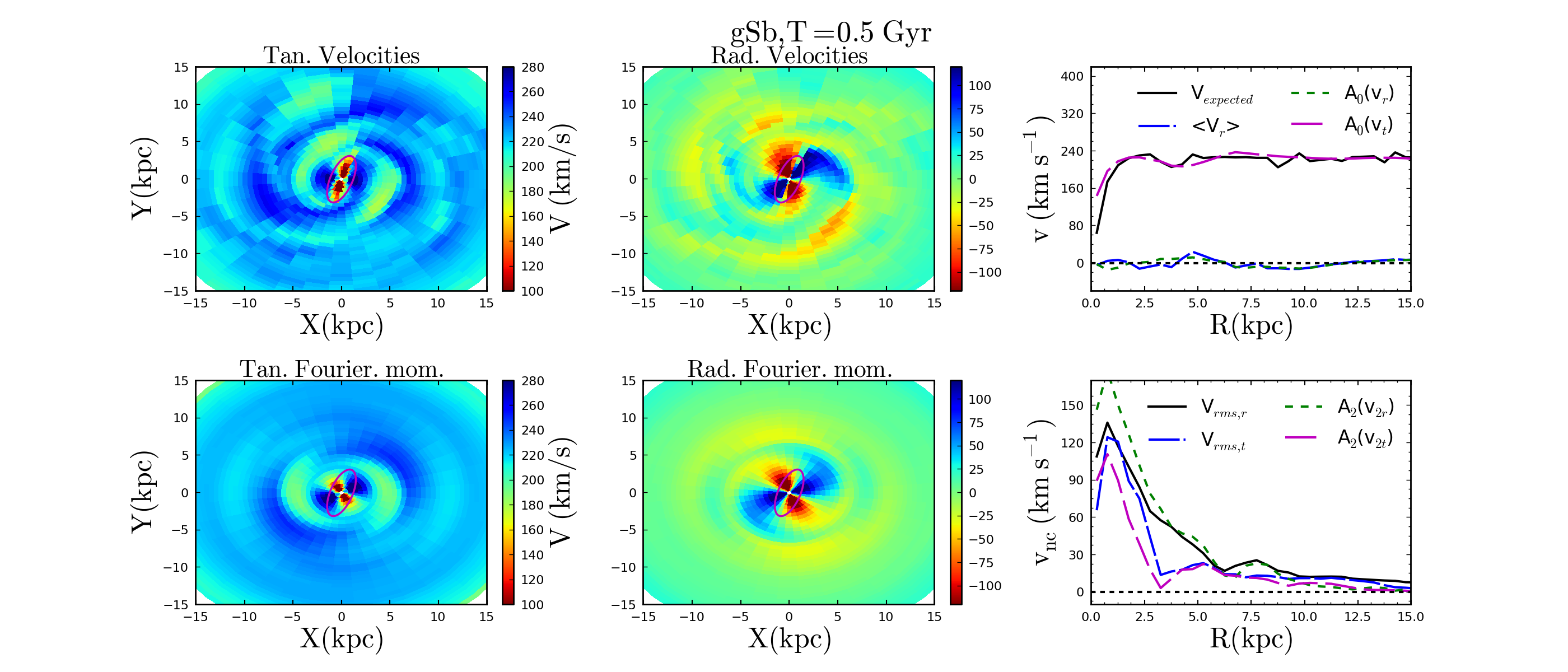}
    \end{array}$
    \end{center}
  \caption{Example of the "intrinsic" rotation curve calculation for the gSb model. { \bf First column:} the tangential velocities map is shown in the top panel and the tangential Fourier moments map in the bottom panel. {\bf Second column}: the radial velocities map is shown in the top panel and the radial Fourier moment map in the bottom panel. {\bf Third column}: the top panel shows the comparison between the expected circular velocities V$_{expected}$ calculated from the gravitational potential as a solid black line and the Fourier A$_{0}$(V$_{t}$) as a long-dashed magenta line. The expected and Fourier axisymmetric radial velocities are shown as solid blue and dashed green lines, respectively. The non-circular velocities as a function of radius are plotted in the bottom panel. The continuous black line and the long-dashed blue line are the rms velocities estimated from the residual maps, and the dashed lines are the m=2 Fourier components. The ellipse in the center of each map shows the location and orientation of the bar.}
 \label{fig4}
\end{figure*}

A sample of the velocity flows found in the simulations is shown in Fig. \ref{fig4}.  The upper left and middle panels show the binned velocity flows for the tangential and radial velocities respectively and the lower left and lower middle panels show the corresponding Fourier models.  The upper right panel shows the expected rotation curve and compares it to the average tangential velocity flow 
and the $m=0$ Fourier moment.  This panel shows that the average velocity, the Fourier moment, and the expected rotation curve are all in agreement.
The lower right panel shows the non-circular radial and tangential velocities calculated either using the rms method or the Fourier moments.
In this case, the Fourier measure is slightly larger than the rms measurement.  This result is easily understood through the method used.  The $m=2$ Fourier moments should equal the maximum deviation from circular motion, while the rms measure should be smaller as it averages the square of the deviations throughout the ring.  

\subsubsection{\rm ROTCUR analysis}
\label{rotc}

\begin{figure}
 \begin{center}$
  \begin{array}{cc}
  \includegraphics[width=85mm]{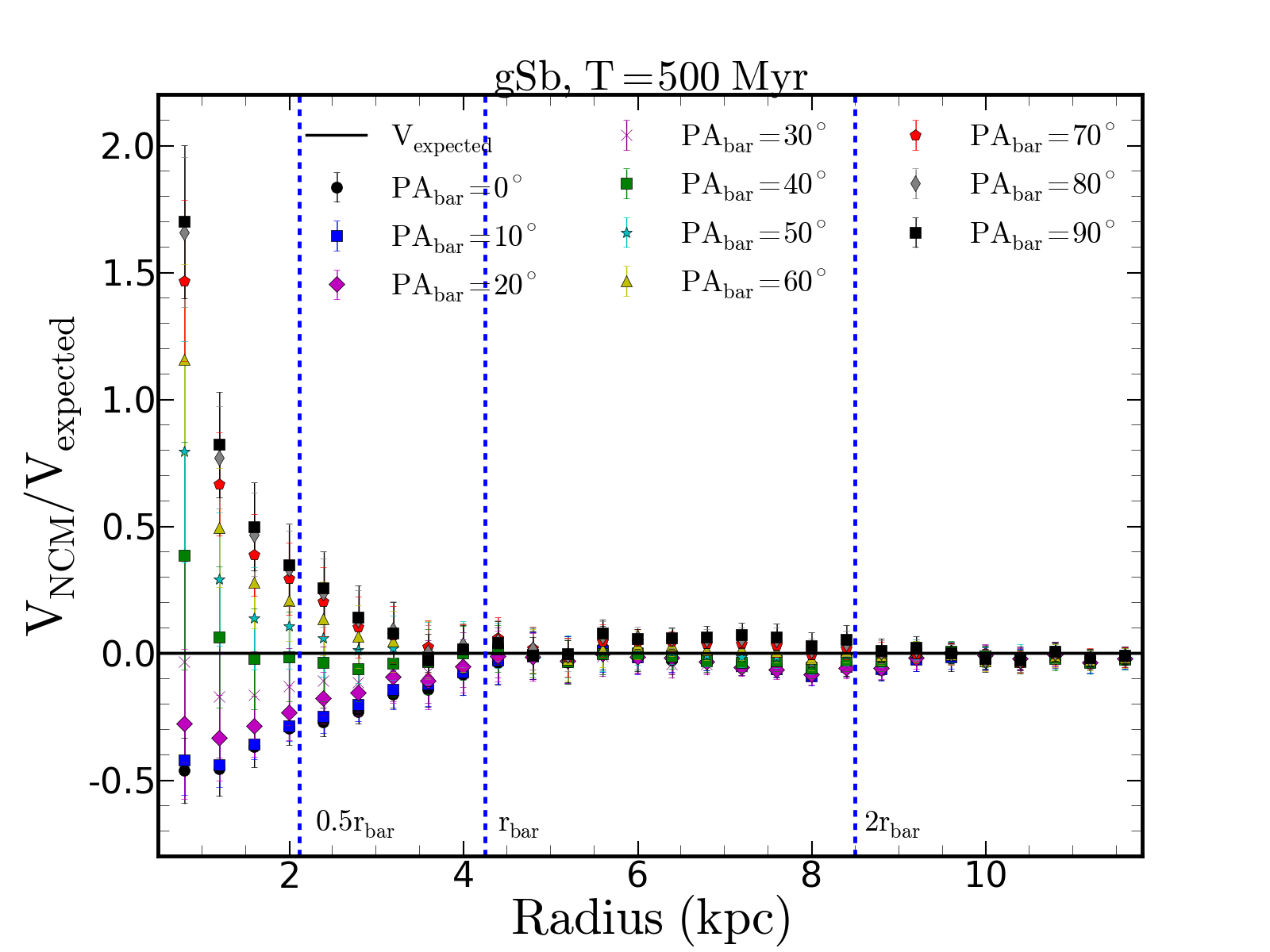}\\
   \includegraphics[width=85mm]{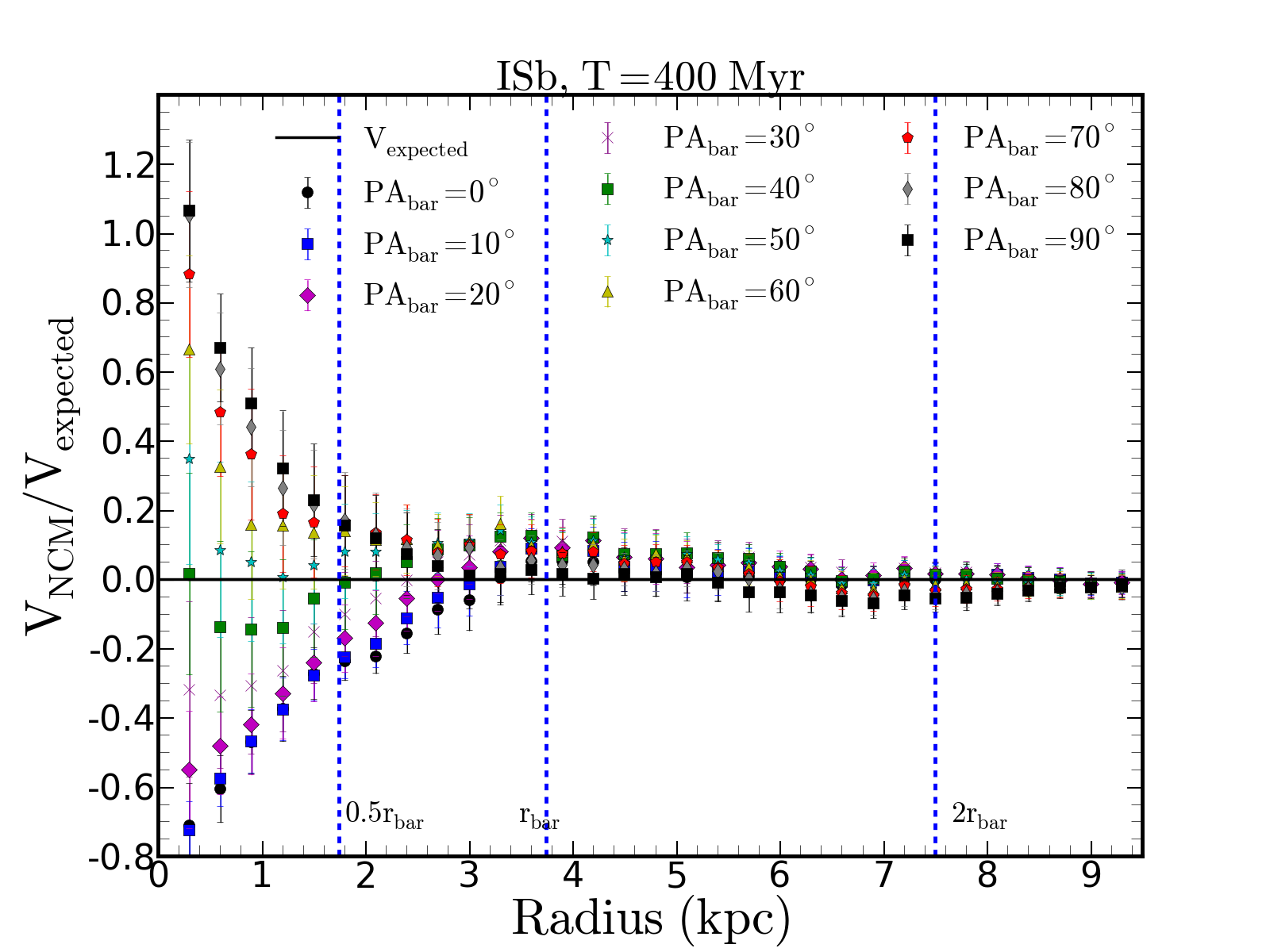}\\
   \includegraphics[width=85mm]{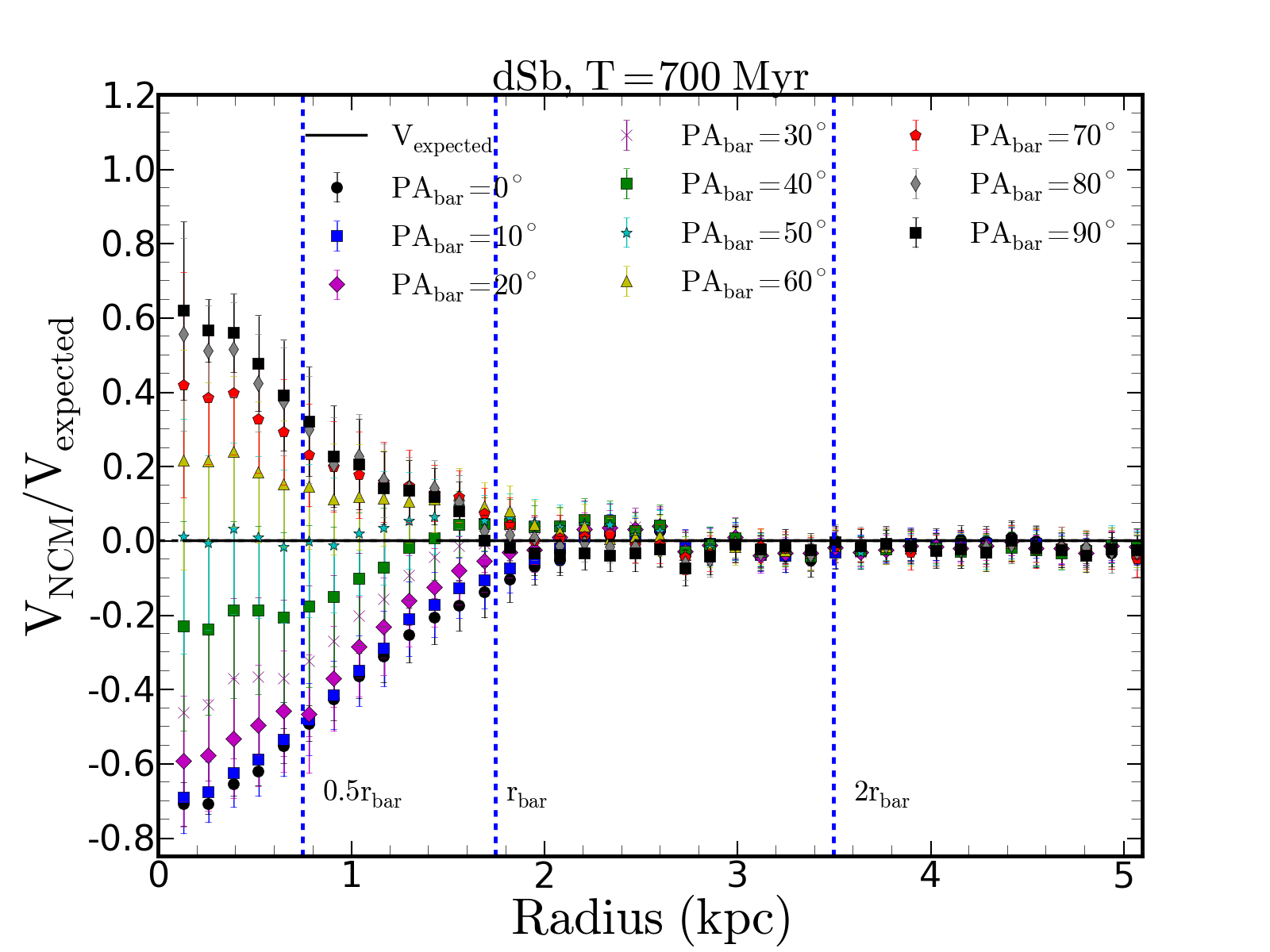}
    \end{array}$
    \end{center}
  \caption{This plot shows the ratio between V$_{NCM}$ = V$_{ROTCUR}$ - V$_{expected}$  and V$_{expected}$  as a function of radius for the gSb, iSb and dSb models.  The vertical dashed lines show 0.5, 1 and 2r$_{bar}$ estimated from the Fourier decomposition. Each rotation curve was derived from mock galaxies with projected bar orientation between 0 and 90 degrees. The projected bar PA are shown on top of each figure.}
 \label{fig5}
\end{figure}

\begin{figure}
 \begin{center}$
  \begin{array}{cc}
   \includegraphics[width=85mm]{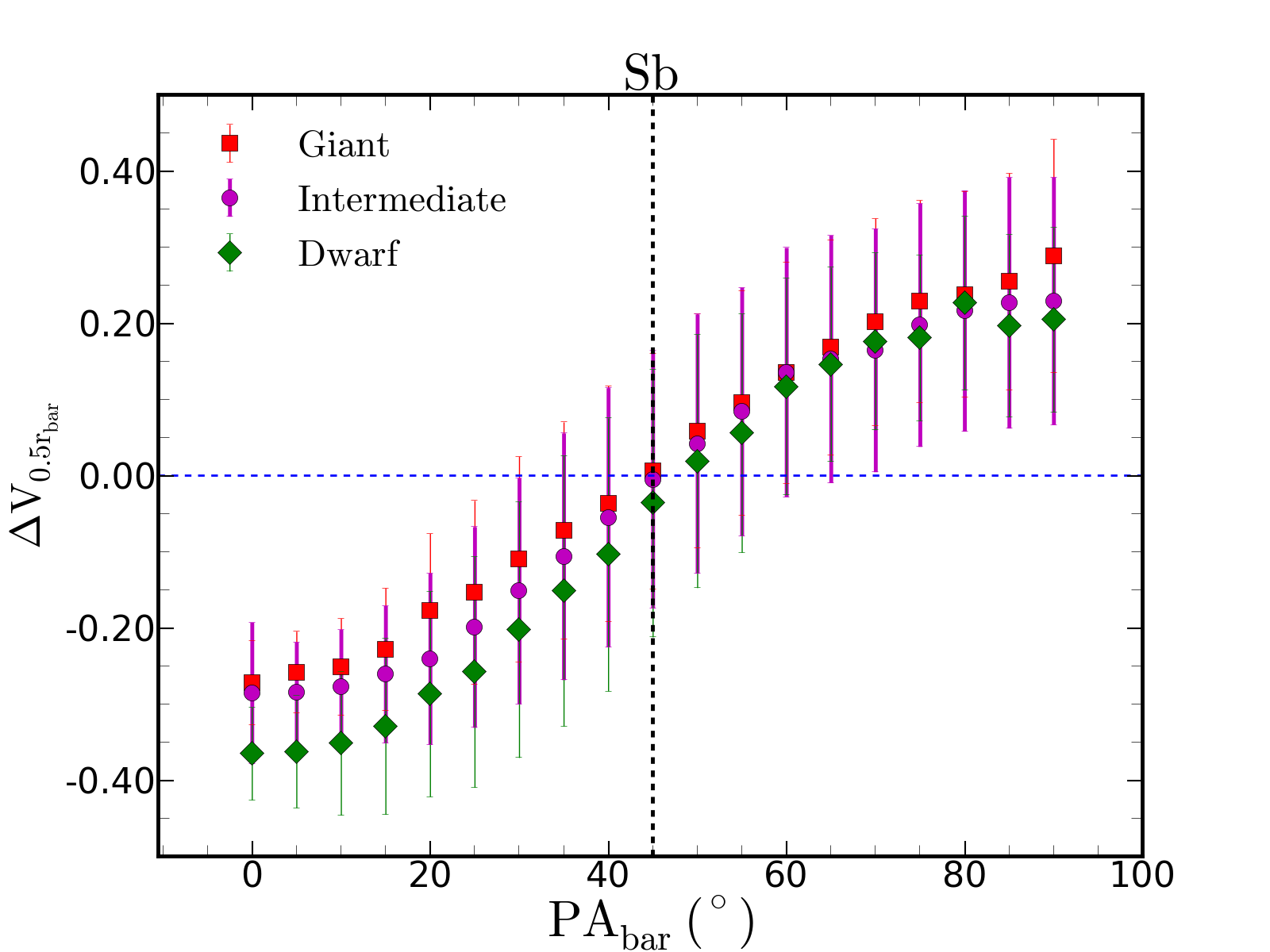}\\
    \end{array}$
    \end{center}
  \caption{Rotation curve error $ \rm \Delta V = [V_{rotcur} - V_{expected}]/V_{expected}$  at R=0.5r$_{bar}$ where r$_{bar}$ is the bar radius estimated from the Fourier decomposition for all the Sb models. The giant Sb model is presented as red squares, the intermediate Sb as magenta circles and the dwarf or low mass Sb as green diamonds.}
 \label{fig6}
\end{figure}

The rotation curves are derived from the velocity maps downloaded from the GalMer database. The bar orientation angle was varied between $0^{\circ}$ and $90^{\circ}$ for each snapshot.  Fig. \ref{fig5} shows the deviations of the ROTCUR rotation curves from the expected rotation curve as a function of radius for the snapshots with the strongest bar for the gSb, iSb and dSb models. The ROTCUR rotation curves agree with the expected curve at $\sim1$ bar length, however they slightly increase between 1 and 2 bar lengths due to the perturbation caused by the spiral arm structures before returning to the true rotation curve at 2 bar lengths. Beyond $\sim2$ bar lengths, the curves all agree with the expected curve. The projected bar PA is shown in the top corner of each panel. The vertical dashed lines show 0.5, 1 and 2 bar radius. The  Fig. \ref{fig5} implies that the rotation curves measured using ROTCUR are smaller than the expected rotation curves inside the bar radius when the bar PA is less than 40$^{\circ}$ and larger than the expected velocities when the projected bar orientation is larger than 60$^{\circ}$. 
\newpage
Fig. \ref{fig6} shows the variation of the rotation curve error, which is the difference between the ROTCUR rotation curve and the expected rotation curve normalized by the expected rotation curve as a function of radius for nine different bar orientations. 
The velocity at R=0.5r$_{bar}$ is used to estimate the rotation curve error. This often corresponds to the maximum ellipticity $\epsilon_{max}$\citep{2006A&A...452...97M}. 
The magnitude of the rotation curve error is zero or very small when the bar is oriented at  45$^{\circ}$ from the major axis. The velocity at half the bar length is more than 40\% larger than the expected value when the bar is perpendicular to the major axis, and 40\% smaller when the bar is parallel to the major axis. There seems to be a systematic increase or decrease of the velocities derived by ROTCUR at half the bar radius depending on the orientation of the bar. This has been previously investigated in the literature (see e.g. \citealt{2007ApJ...664..204S}). The reason for ROTCUR under/over-estimating the velocities is that when the bar is aligned with the major axis, most of the gas and stars are found at  the apocentre of their elliptical orbits at each radius near the major axis. This gas is moving more slowly than it would if it was moving in circular orbits, and the rotation curve is under-estimated if this effect isn't taken into account. Conversely, bars that project along the minor axis place gas and stars close to the pericentre along the major axis, which implies that the rotation curve will be over-estimated if bar-like flows are ignored \citep{2007ApJ...664..204S}. This shows the importance of using a specifically designed packages such as DiskFit when deriving the rotation curves of barred galaxies.
 \subsubsection{\rm DiskFit analysis}
 \label{disk}
 DiskFit \citep{2007ApJ...664..204S} is specifically designed to account for non-circular motions induced by bars.  However, it is well known that the DiskFit algorithm fails when the bar is parallel or perpendicular to the major axis of the galaxy (\citealt{2010MNRAS.404.1733S, 2015MNRAS.454.3743R}). 
 This failure is due to a degeneracy in  equation (10) for $\theta_{b}$ = 0 or 90deg.
\newpage
 The inclination, disk PA, disk kinematic center (xc, yc) and bar PA were allowed to vary during the fits. The systemic velocity $V_{sys}$ is fixed to zero. A bi-symmetric distortion model (m=2) is considered up to R=2.5r$_{bar}$, then the non-axisymmetric flows are set to zero beyond this radius. The disk inclinations and PAs derived from DiskFit are in agreement with the expected values within the error bars. 
   Five bootstraps were used to estimate the uncertainties on the parameters.

\begin{figure*}
 \begin{center}$
  \begin{array}{cc}
  \includegraphics[width=170mm]{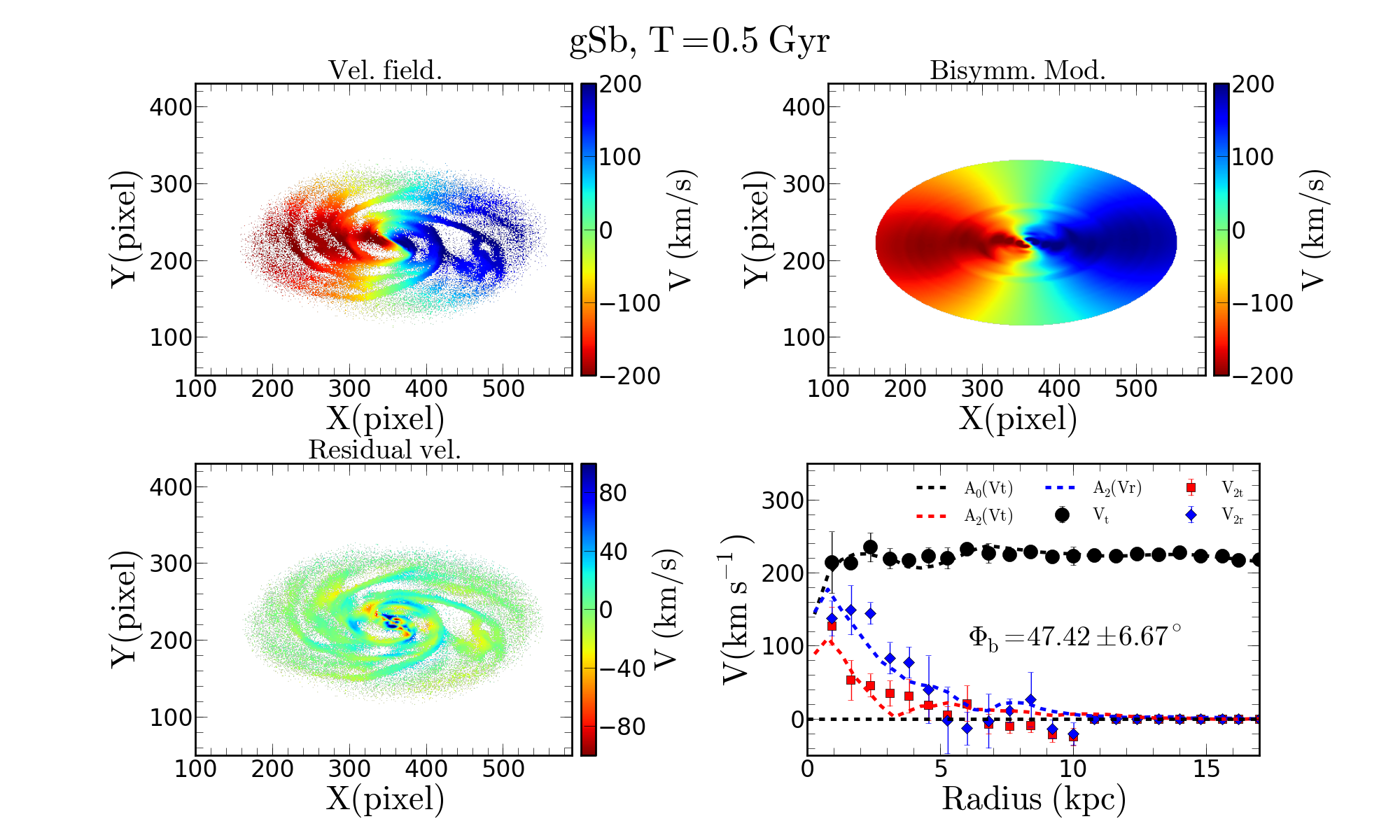}\\
    \end{array}$
    \end{center}
  \caption{DiskFit results for a giant Sb model, Left panel: the observed velocities field (top) and the residual velocities (bottom). Right panel: the DiskFit model (top) and a comparison between V$_{t}$, V$_{2t}$ and V$_{2r}$ with the amplitude of the m=0 and m=2 Fourier mode $A_{0}$ and $A_{2}$ (bottom).}
 \label{fig8}
\end{figure*}

Fig. \ref{fig8} shows a sample DiskFit analysis for the gSb snapshot with a bar angle of $\sim$ $45^{\circ}$. The observed velocity field is in the top left panel,  the DiskFit bi-symmetric model is shown in the top right panel and the residual velocity (data-model) is in the bottom left panel. Only a few pixels in the innermost part of the residual map have values larger than 40 \kms.  

\begin{figure}
 \begin{center}$
  \begin{array}{cc}
  \includegraphics[width=90mm]{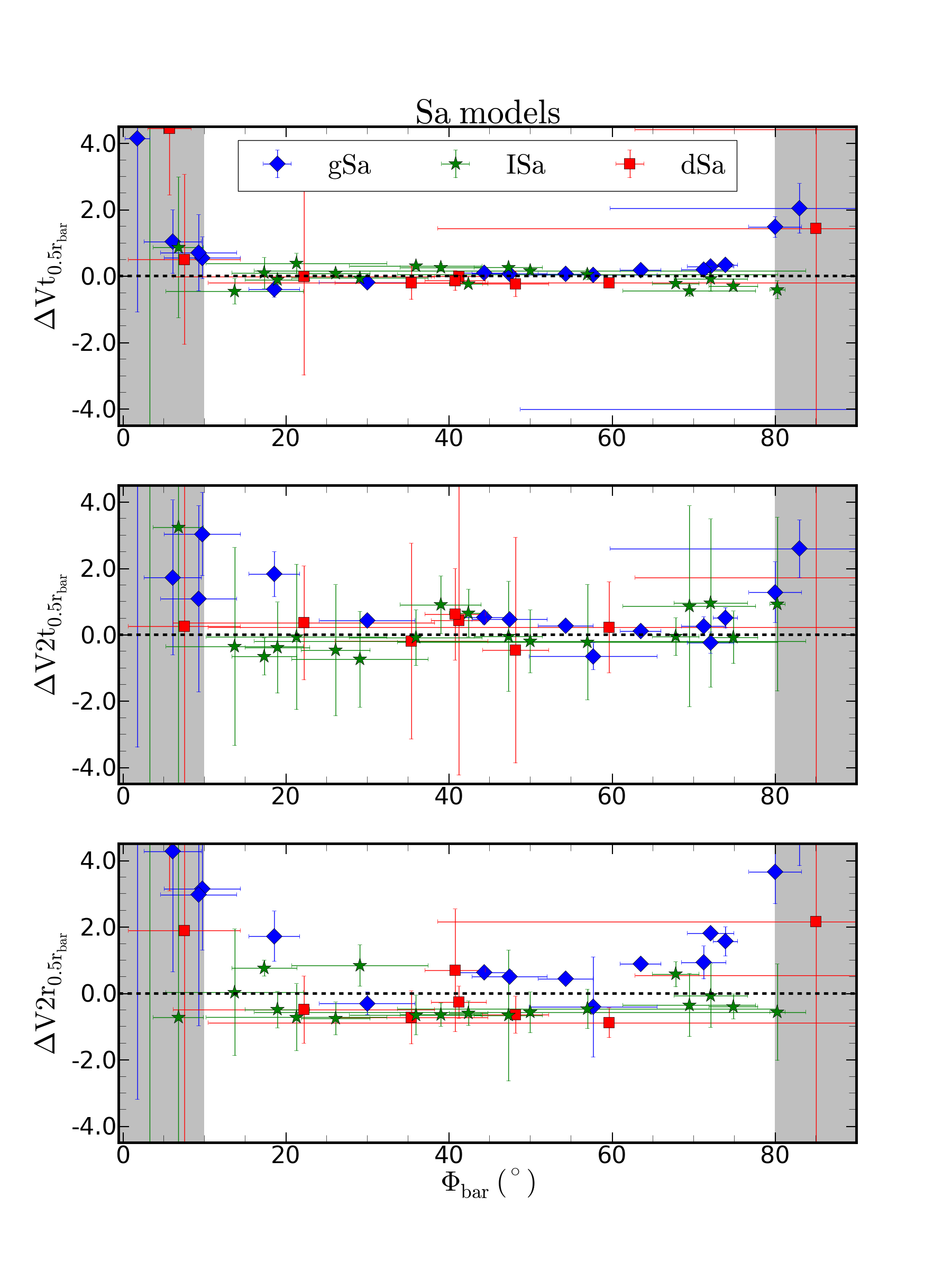}\\
    \end{array}$
    \end{center}
  \caption{Ratio between the Fourier amplitudes and the DiskFit velocities at R=0.5r$_{bar}$ (estimated from the Fourier decomposition of the stellar surface densities) for the spiral Sa models,  giants (blue), intermediate (green) and dwarf (red). The ratio $ \rm \Delta V_{t} = (V_{t} - A_{0,t})/A_{0,t}$ is plotted in the top panel, $ \rm \Delta V_{2r} = (V_{2r} - A_{2,r})/A_{2,r}$  in the middle panel and $ \rm \Delta V_{2t} = (V_{2t} - A_{2,t})/A_{2,t}$ in the bottom panel. The shaded area shows the range of bar orientations where DiskFit fails to recover the bar orientation or velocities, or where the uncertainties on the returned parameters are large.}
 \label{fig9}
\end{figure}

\begin{figure}
 \begin{center}$
  \begin{array}{cc}
  \includegraphics[width=90mm]{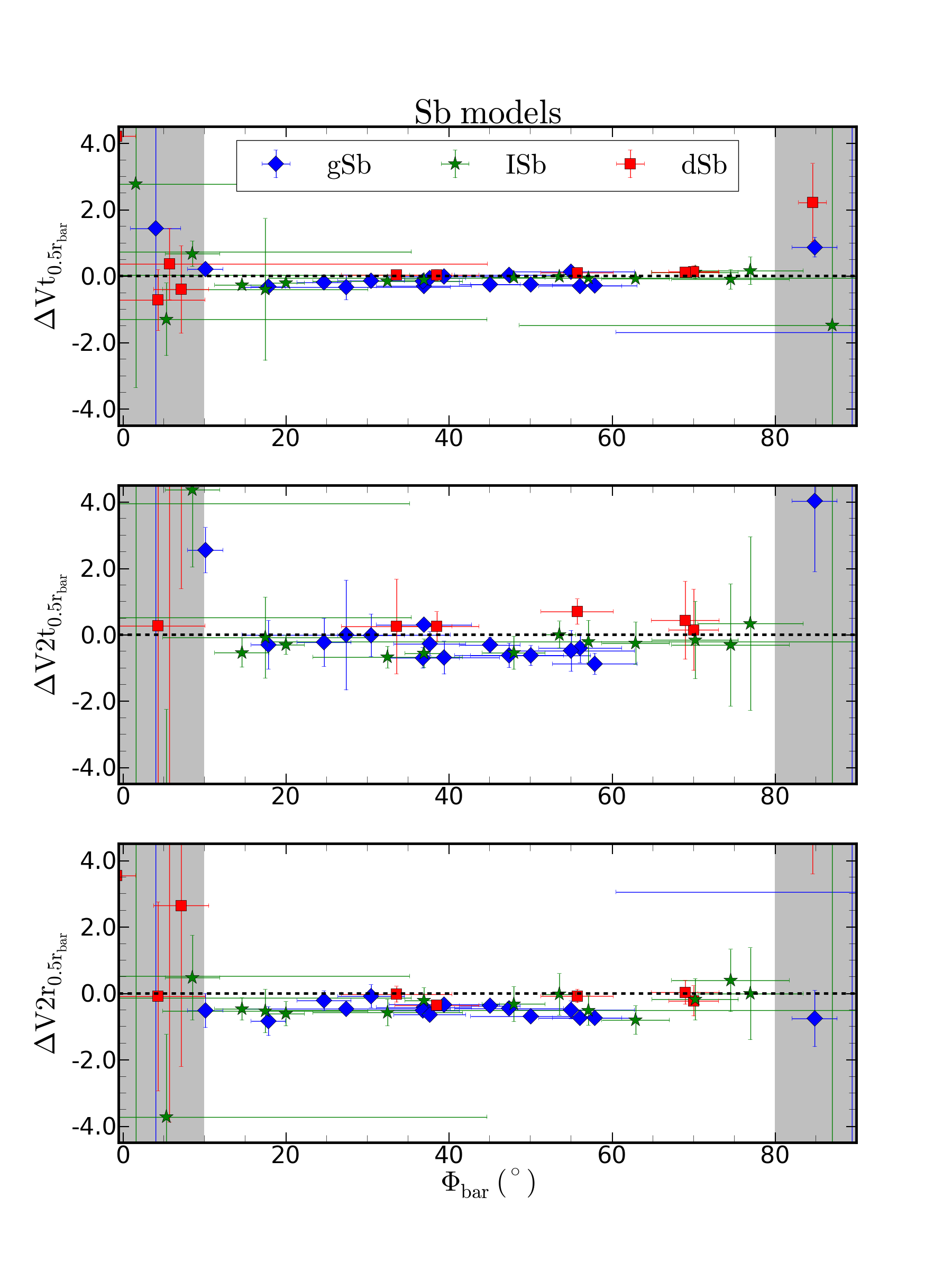}\\
    \end{array}$
    \end{center}
  \caption{Same as Fig \ref{fig9} but for the Sb models.}
 \label{fig10}
\end{figure}

\begin{figure}
 \begin{center}$
  \begin{array}{cc}
  \includegraphics[width=90mm]{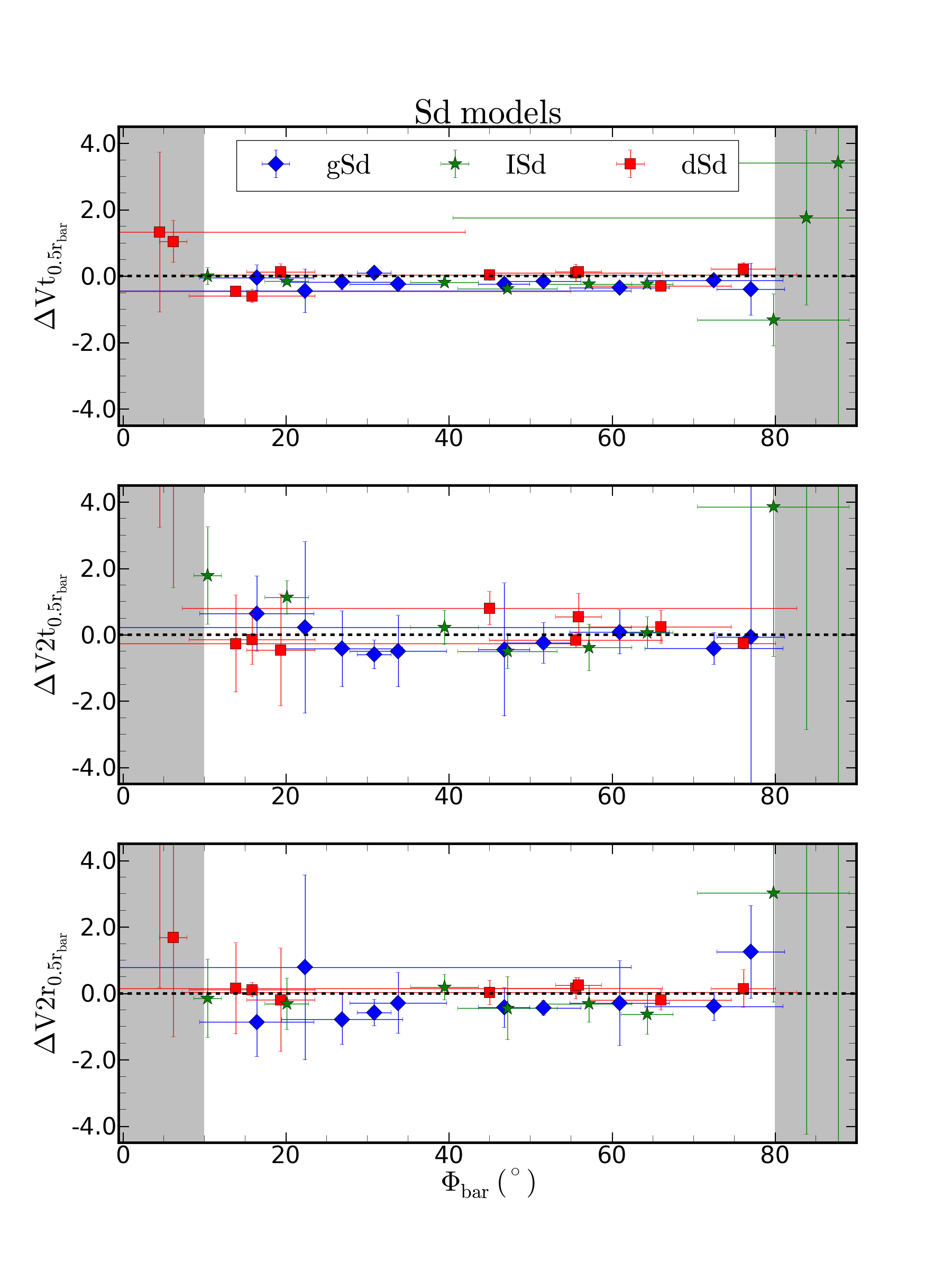}\\
    \end{array}$
    \end{center}
  \caption{Same as Fig \ref{fig9} but for the Sd models.}
 \label{fig11}
\end{figure}

Comparisons between the Fourier amplitudes and the DiskFit velocities for the Sa, Sb and Sd models are shown in Figs. \ref{fig9}, \ref{fig10} and \ref{fig11} respectively. The ratio $ \rm \Delta V_{t} = (V_{t} - A_{0,t})/A_{0,t}$ is plotted in the top panel, $ \rm \Delta V_{2r} = (V_{2r} - A_{2,r})/A_{2,r}$  in the middle panel and $ \rm \Delta V_{2t} = (V_{2t} - A_{2,t})/A_{2,t}$ in the bottom panel. The shaded grey area shows where DiskFit fails to recover the bar orientation or velocities, or where the uncertainties on the returned parameters are large, which is about 10$^{\circ}$ from the symmetry axes. 
There are a number of interesting things to note in these figures. Firstly, DiskFit successfully recovers the rotation curve and the non-circular components for all the models for the bar orientations ranging between $ \sim 10^{\circ}$ and $ \sim 80^{\circ}$. Unlike the ROTCUR analysis, there is no systematic under or over estimation of the velocities. There is a larger scatter in the inferred non-circular moments.  However, the scatter may be due to the resolution of the velocity maps and more work is required before any definite conclusions are reached.

\section{Summary}
\label{sum}

We present an analysis of the strongest barred snapshots of nine isolated GalMer simulations.  We 
characterized the intrinsic bar strengths and non-circular motions and compared them to mock observations of the systems.  

We found that:
\begin{enumerate}
  \item The bar lengths estimated from ellipse fitting are shorter than those estimated from the Fourier decomposition technique which is consistent with  previous studies (e.g. \citealt{2005MNRAS.364..283E, 2009A&A...495..491A}).
   \item The Fourier decomposition technique gives similar results for the bar properties when applied directly to the simulation or to the images obtained from mock observations.
   \item ROTCUR systematically under-estimates the velocities when the bar is parallel to the major axis and overestimates them when the bar is perpendicular to the major axis for all the models. This shows the importance of properly accounting for the non-circular motions induced by the bar when deriving rotation curves (see \citealt{2007ApJ...664..204S, 2008MNRAS.385..553D, 2015MNRAS.454.3743R}).
  \item  The magnitude of the rotation curves error when ROTCUR is used depends only on the orientation of the bar and it is not correlated with either the bar strength or the bar length.
   \item  DiskFit is able to reproduce the amplitudes of the Fourier m=0 and m=2 components when the bar is at an intermediate PA. 
    \item Unlike ROTCUR, DiskFit does not show a systematic over-estimation or under-estimation of the inner part of the rotation curves.
  \item  DiskFit fails to recover the bar orientation or velocities, or return large parameter uncertainties when the bar is closer than 10 degrees to either the minor or the major axis (see also \citealt{2010MNRAS.404.1733S}).
\end{enumerate}
As future work, we will be using new simulations of a sample of four barred galaxies. This will allow us to perform a direct comparison between the simulation with the HI and ancillary data from observations, as well as putting a tighter constraint on the bar orientation range where DiskFit fails.

\section*{Acknowledgments}
We thank the anonymous referee for the comments and suggestions. 
CC's work is based upon research supported by the South African Research Chairs Initiative (SARChI) of the Department of Science and Technology (DST), the SKA SA and the National Research Foundation (NRF). ND and TR's work is supported by a SARChI's South African SKA Fellowship. KS acknowledges support from the National Sciences and Engineering Research Council of Canada (NSERC).

\bibliographystyle{aa}

\newpage

\end{document}